\documentclass[twocolumn,floatfix]{revtex4}
\usepackage{graphicx}

 \newcommand{\beq}    {\begin{equation}}
 \newcommand{\eeq}    {\end{equation}}
\begin{document}

\title{Nonlinear spectroscopy of photons bound to one atom}

\author{I. Schuster, A. Kubanek, A. Fuhrmanek,
T. Puppe, P.W.H. Pinkse, K. Murr \& G. Rempe}
\email{gerhard.rempe@mpq.mpg.de}
\affiliation{ Max-Planck-Institut f\"ur Quantenoptik, Hans-Kopfermann-Str. 1,
D-85748 Garching, Germany.}
\normalsize
\begin{abstract}
Optical nonlinearities typically require macroscopic media, thereby making
their implementation at the quantum level an outstanding challenge.
Here we demonstrate a nonlinearity for one atom enclosed by two
highly reflecting mirrors\cite{Berman94}. We send laser light through
the input mirror and record the light from the output mirror of the
cavity. For weak laser intensity, we find the vacuum-Rabi
resonances\cite{Boca04,Maunz05,Puppe07,Wallraff04,
Reithmaier04,Yoshie04,Peter05,Khitrova06,Press07,Hennessy07}.
But for higher intensities, we find an additional
resonance\cite{Carmichael94}.
It originates from the fact that the
cavity can accommodate only an integer number of photons and that
this photon number determines the characteristic frequencies of the
coupled atom-cavity system\cite{Rempe87,Brune96,Schuster07}.
We selectively excite such a frequency by
depositing at once two photons into the system and find a transmission
which increases with the laser intensity squared. The nonlinearity
differs from classical saturation
nonlinearities\cite{Rempe91,Lugiato92,Srinivasan07,Englund07} and is
direct spectroscopic proof of the quantum nature of the atom-cavity
system. It provides a photon-photon interaction by
means of one atom, and constitutes a step towards a two-photon gateway
or a single-photon transistor\cite{Chang07}.
\end{abstract}
\maketitle
\begin{figure}
\includegraphics[width=\columnwidth]{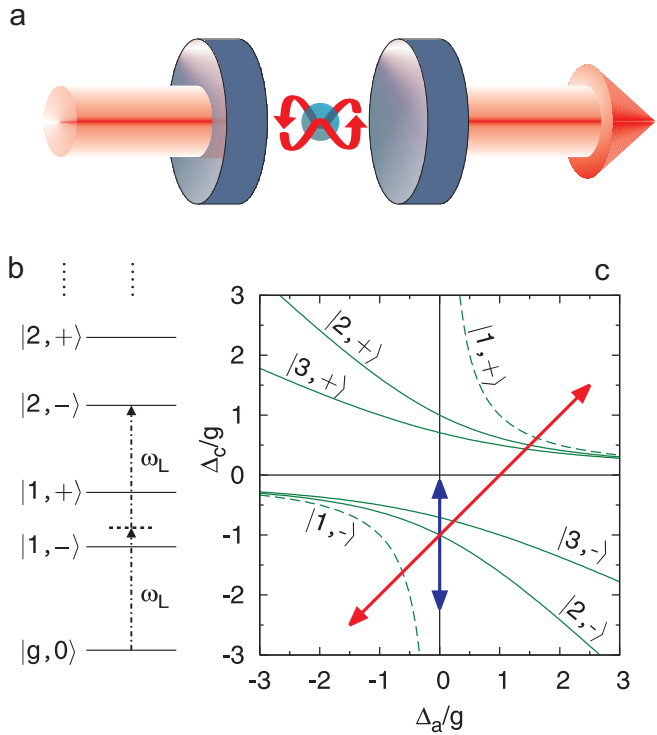}
\caption{\label{Figure1}{\bf The atom-cavity molecule.} {\bf (a)}
Artist's conception of an atom enclosed by two cavity mirrors. The
coupling of the atom to the intracavity light is so strong that the
energy spectrum of the composite atom-cavity `molecule' radically
differs from the spectrum of its components. {\bf (b)} Schematic of
such a level structure illustrating how the dressed state
$|2,-\rangle$ is directly accessed from the vacuum via a two-photon
transition.
{\bf (c)} Location of the classical (normal-mode) resonances (dashed
curves) and the first two quantum resonances (states $|2,\mp\rangle$
and $|3,\mp\rangle$) (solid curves). The diagonal arrow is the scan
direction shown in Fig.~\ref{Figure2}, which crosses all resonances.
A vertical scan along
$\Delta_a=0$, as in Fig.~\ref{Figure3}, avoids the classical
resonances. Resonances for the manifold ($n+1$) are located at
$\Delta_c=\pm g/\sqrt n$ for $\Delta_a=0$.}
\end{figure}

The quantum nonlinearity has its origin in the fact that under the
condition of strong coupling a system composed of a single atom and
a single cavity mode has properties which are distinctively
different from those of the bare atom (without the cavity), or the
bare cavity (without the atom), or just the sum of the two
(Fig.~\ref{Figure1}a). In fact the composite system forms
a new quantum entity, the so-called atom-cavity molecule, made of
matter and light, with its own characteristic energy spectrum. This
spectrum consists of an infinite ladder of pairs of states, the dressed
states (Fig.~\ref{Figure1}b). The first doublet contains one quantum of
energy and can be probed by laser spectroscopy. For weak probing,
the resulting spectrum is independent of the laser intensity and has
been dubbed the vacuum-Rabi or normal-mode spectrum, consisting of a
pair of resonances symmetrically split around the bare atomic and
cavity resonances. This spectrum has first been observed with atomic
beams\cite{Berman94}, and has been explored recently with single
dipole-trapped atoms\cite{Boca04,Maunz05,Puppe07}.
It constitutes a benchmark for strong
atom-cavity coupling and is central to most cavity quantum
electrodynamics (QED) experiments, including those outside atomic
physics\cite{Wallraff04,Reithmaier04,Yoshie04,Peter05,
Khitrova06,Press07,Hennessy07}.
Notice that the normal-mode spectrum on its own can equally well be
described classically, by linear dispersion theory or a
coupled oscillator model (atomic dipole and cavity field). \par
The next higher lying doublet contains two quanta of energy
and lacks a classical
explanation\cite{Carmichael94,Carmichael96,Thompson98}.
The corresponding dressed states have been observed
(together with a few higher order states) in
microwave cavity QED\cite{Rempe87,Brune96,Schuster07}
and even ion trapping, where phonons play the role of
photons\cite{Meekhof96}. At optical frequencies, evidence for these
states has indirectly been obtained in
two-photon correlation experiments where the conditional response of
the system upon detection of an emitted photon is
monitored\cite{Rempe91,Mielke98,Foster00,Birnbaum05}. These optical
experiments observe the quantum fluctuations in dissipative cavity
QED systems but operate away from a resonance to a
higher-lying state.\par
Our experiment exploits the anharmonicity of the energy-level
spectrum to drive a multi-photon
transition directly from the vacuum state to a specific higher lying
state. We observe the quantum character of our
cavity QED field by measuring a photon flux, not a photon
correlation.
To explain our technique, we note that a two-state
atom coupled to a single-mode light field has a discrete spectrum
consisting of a ladder of dressed states, $|n+1,\mp\rangle$, with
frequencies
\beq
\label{QuantumFrequencies}
\omega_{n+1,\mp}\!=n\omega_c\!+\frac{1}{2}(\omega_a+\omega_c)
\mp\frac{1}{2}\sqrt{4g^2(n\!+1)+(\omega_a\!-\omega_c)^2}
\eeq
and a ground state $|g,0\rangle$ with zero energy (atom with ground
state $|g\rangle$ and mode in the vacuum state $|0\rangle$). Here,
$n=0, 1, 2, \dots$ is the principal quantum number of the mode (to
be distinguished from the mean photon number), $g$ the atom-cavity
coupling strength, and $\omega_a$ and $\omega_c$ are the frequencies
of the atom and the cavity, respectively. The frequencies of the
coupled system are probed with monochromatic light of frequency
$\omega_L$\cite{Carmichael94}. Resonances occur when
$(n+1)\omega_L=\omega_{n+1,\mp}$. In Fig.~\ref{Figure1}c, these
resonance conditions are plotted in the frame ($\Delta_a,\Delta_c$)
where $\Delta_a=\omega_L-\omega_a$ and $\Delta_c=\omega_L-\omega_c$
are the atom and cavity detunings, respectively.\par
We now notice that if each laser photon is resonant with the atom,
$\omega_L=\omega_a$ ($\Delta_a=0$), it is detuned from the
single-photon resonances, $\omega_L\neq\omega_{1,\mp}$, and this for
\emph{any} value of the cavity frequency $\omega_c$ (vertical arrow
in Fig.~\ref{Figure1}c). Scanning the cavity frequency around the
frequency of a weak laser therefore gives a suppressed and largely
frequency-independent response, as further discussed in connection
with Fig.~\ref{Figure3}. When increasing the laser intensity,
however, two-photon transitions can occur at $\Delta_c=\mp g$, where
two laser photons together are resonantly absorbed by the combined
system and the second manifold of dressed states is populated for
$2\omega_L=\omega_{2\mp}$.
We keep the intensity at low enough values such that the
atomic transition is never saturated. In this way, we rule out the
possibility of nonlinearities due to a classical behavior of the
intracavity field\cite{Lugiato92}.
This protocol of driving multi-photon
transitions by avoiding the normal modes as well as atomic saturation
is new and should in
principle apply to other cavity QED systems. It can be interpreted
as a two-photon gateway: single photons cannot be accepted by the
combined atom-cavity system, but two photons can.\par
%
Our implementation of a strongly-coupled atom-cavity system consists
of single ${}^{85}$Rb atoms localized inside the mode of a
high-finesse optical cavity by means of an auxiliary intracavity
red-detuned dipole trap at 785 nm. We do spectroscopy on the system
by shining near-resonant probe light at 780 nm onto the input mirror
and recording the light exiting from the output mirror.
While probing the system,
we also monitor the localization of the atom. We then postselect
only the events for which the condition of strong coupling was
fulfilled. For sufficient statistics,
we average over many trapping events
(see Appendix \ref{App::Methods}).\par
\begin{figure}
\includegraphics[width=\columnwidth]{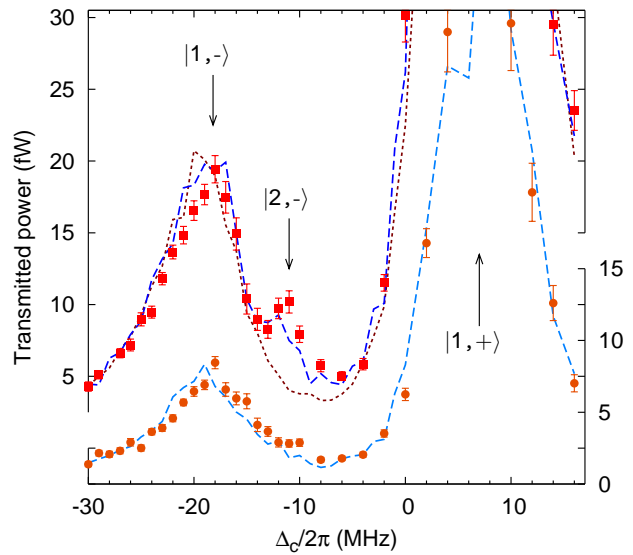}
\caption{\label{Figure2}{\bf Quantum anharmonicity of the
atom-cavity system.} Transmitted power as a function of cavity
detuning $\Delta_c$ for the input powers $0.5$~pW (circles, right
scale) and $1.5$~pW (squares, left scale). In this and all following
figures, $10$ fW correspond to about $0.01$ intracavity photons, the
error bars are s.d. and an offset of $0.5$ fW due to detector dark
counts has been removed from the data. The two peaks at
$2\pi\times(-18,7)$ MHz represent the two normal modes,
which are reproduced
by a theory considering only single-photon transitions (dotted line).
Note that the maximum of the right peak is beyond the scale of the
figure. The quantization of the intracavity light (dashed lines) is
required to explain the appearance of the two-photon
resonance at $-2\pi\times11$ MHz at higher intensity.}
\end{figure}
\begin{figure}
\includegraphics[width=\columnwidth]{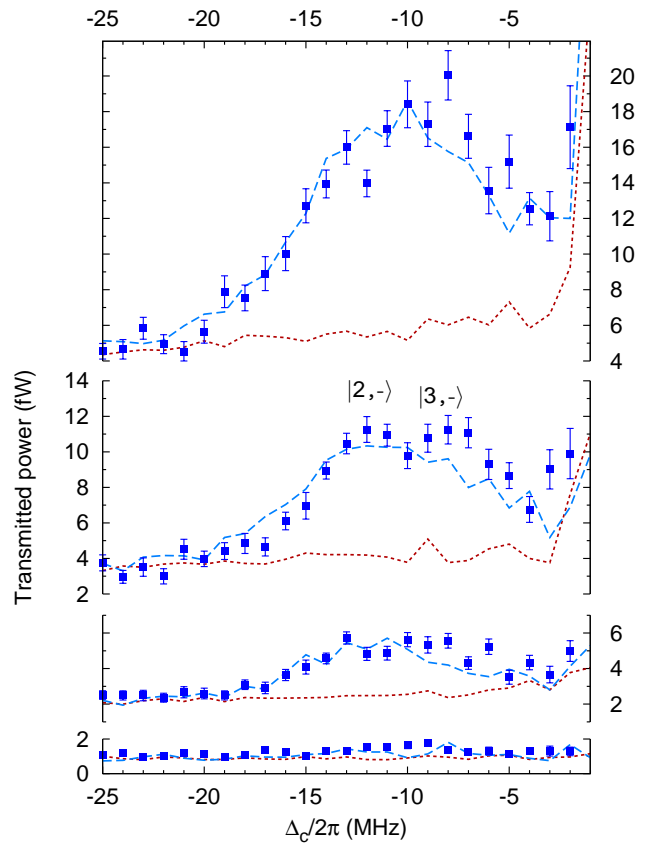}
\caption{\label{Figure3}{\bf Multi-photon resonances and suppression
of single-photon resonances.} Transmitted power as a function of
cavity detuning $\Delta_c$ for input powers of ($0.5,1.5,2.4,3.3)$
pW (from bottom to top, data points). Also shown are the simulations
with single-photon (dotted lines) and multi-photon transitions
(dashed lines), with the respective positions of the dressed state
$|2,-\rangle$ and $|3,-\rangle$.}
\end{figure}

In a first experiment, the cavity transmission is monitored for an
atom-cavity detuning $\Delta_a-\Delta_c \approx g$. For these
parameters, the normal-mode spectrum becomes asymmetric and the
splitting between states $|1,-\rangle$ and $|2,-\rangle$ increases
compared to the case $\Delta_a \approx \Delta_c$. This has the
advantage that this two-photon resonance is well separated
from the normal modes even for moderate atom-cavity coupling
constants. Fig.~\ref{Figure2} displays two scans with different
probe laser intensities along the diagonal arrow in
Fig.~\ref{Figure1}c. Both scans show the normal modes, but the
higher-intensity scan shows a pronounced additional resonance.\par
The quantum anharmonicity\cite{Brune96} of the energy-level
structure reveals itself in the position of the two-photon
resonance relative to the normal modes, which is in excellent
agreement with equation~(\ref{QuantumFrequencies}) for a coupling of
$g=2\pi\times11.2$ MHz and a Stark-shifted atom-cavity detuning of
$2\pi\times10.5$ MHz. These values indicate that the coupling is about
70\% of the maximally possible value
$g_0=2\pi\times16$ MHz at an antinode, and
that the trap induces an average Stark shift of
$2\pi\times24.5$ MHz which also corresponds to about 70\% of
the Stark shift at an {anti\-node} of the standing wave dipole trap
(the trap depth is 170 nW and the bare atom-cavity detuning is preset
to $2\pi\times35$ MHz). Ideal couplings are not
reached due to two reasons: firstly, the atom performs an oscillation
in the dipole trap wells; secondly, the position of the trapping
wells shifts with respect to the antinodes of the probe light along
the cavity axis as the distance from the cavity center increases,
therefore atoms which are trapped slightly off the cavity center
are not maximally coupled.\par
We next performed extensive numerical simulations in order to compare
the measured spectrum to several cavity QED
models\cite{Carmichael94}.
The simulations closely imitate the experiment, starting from the
trapping of single, slow atoms which are injected into the intracavity
dipole trap, following up with the measurement protocol
which is executed until the atom leaves the trap, and culminating in
the same evaluation procedure (see Appendix \ref{App::Analysis}).
The first set of
simulations (only shown for the higher-intensity scan, dotted line
in Fig.~\ref{Figure2}) assumes at most one quantum of energy in the
system, thereby allowing us to quantify the
contribution due to single-photon transitions.
We see that, while the normal-mode resonances are reproduced,
there is a large deviation to the measured data precisely at the
position of the two-photon resonance.
In contrast, another set of simulations (dashed lines) was
performed for a quantized cavity mode with three Fock states,
$|0\rangle,|1\rangle,|2\rangle$.
Apart from the normal modes, these simulations also
reproduce the two-photon resonance.\par
In a second experiment, we explore the quantum regime by scanning the
cavity along the vertical arrow in Fig.~\ref{Figure1}c, using a bare
atom-probe detuning of $2\pi\times21$~MHz and dipole trap power of 140
nW. Fig.~\ref{Figure3} displays four scans with different probe laser
intensities. The lowest intensity scan (bottom panel) shows a largely
flat signal, in agreement with our idea of avoiding the normal modes.
Here, the photonic state is close to the vacuum state, with a mean
intracavity photon number as low as $0.001$. All higher-intensity scans,
however, show a pronounced additional resonance. The deviation between
the off-resonance signal, $\Delta_c/2\pi\leq-20$~MHz, and the
on-resonance signal, $-15$ MHz$\leq\Delta_c/2\pi\leq-5$~MHz, gets larger
with increasing intensity. Here, we find that a simulation with at most
one quantum of excitation (dotted lines) continues to predict a signal
with no major modification in the structure, whereas the resonance is
globally reproduced with simulations taking into account field
quantization. We had to account for four Fock states, indicating that the
resonance stems from a two-photon and a weak but rising three-photon
transition, the latter transition contributing to broaden the resonance.
We also notice an increase in the transmitted intensity at small cavity
detunings $|\Delta_c|/2\pi \leq 3$ MHz. This is a systematic effect
originating from the bare cavity resonance occurring at $\Delta_c=0$ with
a linewidth of $\kappa=2\pi\times1.25$~MHz, which could not be completely
suppressed in the post-selection process (see Appendix
\ref{App::Methods}). \par
\begin{figure}
\includegraphics[width=\columnwidth]{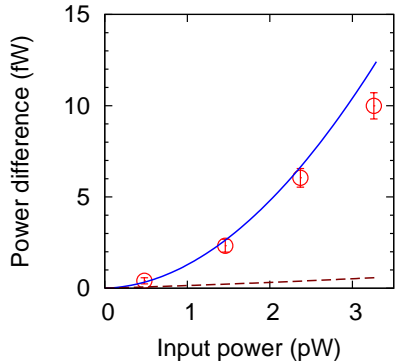}
\caption{\label{Figure4}{\bf Nonlinear intensity response.}
Difference between the nonlinear response on the two-photon resonance
and the linear single-photon response, together with the prediction
from a quantum theory with an immobile atom (solid line).
For reference we also show
the nonlinearity expected from the saturation of a two-state atom
coupled to a classical field (optical bistability theory,
dashed line).}
\end{figure}

Under coherent excitation, we expect a mainly {quad\-ra\-tic} scaling
of the transmitted intensity at the frequency of the two-photon
resonance and a linear scaling off resonance, i.e. at large cavity
detunings $|\Delta_c|\gg g$, where all theories coincide.
To evaluate the quantum response of the system, the data
used for Fig.~\ref{Figure3} are averaged over the two-photon region
($-15$ MHz $\leq\Delta_c/2\pi\leq-10$ MHz) as well as over the
off-resonance region ( $-25$ MHz $\leq\Delta_c/2\pi\leq$ $-20$ MHz),
for each input intensity. The off-resonant region serves as a reference
for the linear single-photon contribution.
By taking the difference between the two regions, we isolate the
contribution of the two-photon transition, and find a mainly quadratic
response of the transmitted intensity versus the input intensity
(Fig.~\ref{Figure4}, circles). \par
We proceed by comparing the data to a model which assumes the atom
to be immobile. This model describes an ideal quantum system,
with a well-defined coupling to the mode, as would be desirable for
future applications.
To this end, we match all the spectra of Fig.~\ref{Figure3} to
this idealized theory with a common set
of parameters $(g,\Delta_a)$, and find good agreement for
$(g,\Delta_a)=2\pi\times(11.5,1)$ MHz (see Fig.~\ref{fig:supp3}).
Notice that the coupling $g$ is close
to the value obtained for the independent measurements in
Fig.~\ref{Figure2}, whereas the atomic detuning $\Delta_a$
is close to zero. The theory spectra are then evaluated in the same
way as the measured spectra to obtain the intensity response of the
system. The resulting nonlinear curve (solid line in Fig.~\ref{Figure4})
describes the measured data reasonably well and also shows a mainly
quadratic dependency. We hope to further approach the
fixed-atom limit in our experiment by extending our cooling
protocol from one to three dimensions\cite{Nussmann05}. \par
For completeness we also analyzed the nonlinear theory of
optical bistability\cite{Rempe91,Lugiato92} and found that it is
inconsistent with all the measurements presented in this paper, as shown
with simulations
(Appendix \ref{App::Analysis}, Figs. \ref{fig:supp1} and \ref{fig:supp2}).
Specifically, the bistability theory predicts a behavior close to
the one we obtained with the theory of coupled oscillators.
Explained differently,
according to bistability theory, we are operating on the lower branch,
where the corresponding nonlinear response is small (dashed line in
Fig.~\ref{Figure4}). Indeed, the reported nonlinearity occurs with
an occupation probability of the atomic excited state of at most
$0.07$. This is what makes it radically different from and dominant
over the standard saturation nonlinearity for a two-state atom. \par
In conclusion, our experiment enters a new regime,
with nonlinear quantum optics at the level of individual atomic and
photonic quanta. In the future we plan to investigate the photon
statistics and the spectrum of the light transmitted
through the cavity. Once improved cooling forces are implemented to
better localize the atom in the cavity mode, new multi-photon states
could be produced by applying techniques originally developed for a
single atom and single photons to the case of a single atom-cavity
system and multiple photons. Other applications in quantum
information science include a single-photon transistor, where one
photon controls the propagation of another photon\cite{Chang07}.\par


\appendix
\section{Experimental setup and measurement procedure \label{App::Methods}}
\subsection{Setup}
The cavity supports a TEM$_{00}$ mode near-resonant with the $5
{}^2S_{1/2} F = 3, m_F = 3 \rightarrow  5 {}^2P_{3/2} F = 4, m_F =4$
transition of ${}^{85}$Rb atoms at wavelength $\lambda =
780.2$ nm. The input power is given as the integrated intensity
measured in transmission of the resonant empty cavity; the average
intracavity photon number is about 0.9/pW (which is approximately
$10$ fW for $0.01$ photon as given in the paper). Our cavity does
not show birefringence. Consequently, the circular polarization of
the input probe light ensures that the atomic transition is closed
so that the atom effectively has only two internal states. In
addition, a far-detuned dipole-trap laser at 785.3 nm resonantly
excites a second TEM$_{00}$ mode supported by the
cavity\cite{Maunz04},
 two free spectral ranges detuned from the
near-resonant probe mode. Antinodes of probe light and dipole trap
coincide in the center of the cavity, such that atoms are localized
at regions of strong coupling. The dipole trap laser also serves to
stabilize the cavity length. Single atoms are injected into the
cavity by means of an atomic fountain
and trapped by switching the power of the dipole-trap laser
as soon as they reach a region of strong coupling.
\subsection{Scanning procedure}
After trapping, single atoms are first observed and
cooled\cite{Maunz04}
with a weak laser with a power of 0.3 pW and a detuning of $\Delta_c$ = 0
for a time interval of $500\ \mathrm{\mu s}$, which we shall call
the check interval. This
light is then switched to higher power and its detuning is
simultaneously adapted in order to probe the system at different
frequencies for $100\ \mathrm{\mu s}$ (hereafter the probe interval).
The switching between low and high power has been optimized for high
speed with minimal overshoots and drifts. This sequence of check and
probe intervals is repeated either 20 or 30 times, depending on the
expected trapping time of the atom. For Fig.~\ref{Figure2}, for each data point a
different probe frequency $\omega_L$ is chosen. For Fig.~\ref{Figure3}, we
vary the cavity length between different trapping events, thereby
tuning the cavity frequency $\omega_c$. During the probe
intervals, the laser is switched to a fixed frequency in the vicinity
of $\Delta_a$ = 0, the Stark-shifted resonance of the uncoupled trapped
atom. We record the photons exiting from the output cavity mirror
during the whole measurement and for all atoms
and then apply a post-selection protocol as described below.
\subsection{Post-selection and accessible scan regions}
Atoms which have been strongly heated during probing
(due to spontaneous emission or cavity-mediated
forces\cite{Hechenblaikner98}) will follow an
enhanced oscillation in the trap, and thus the coupling to the probe
mode will be reduced. In this case the interval needs to be removed
from the data sample. In order to identify such probe intervals, we
check the coupling to the probe mode before and after each probe
interval by observing the transmission on the resonance of the empty
cavity $\Delta_c=0$, which is strongly suppressed in case of a
well-localized atom. In a
postselection protocol, we select only those probe intervals which
show good localization in both enclosing check intervals
for further evaluation. About 15\% of all intervals
in which an atom was present survive this procedure if we require a
minimal coupling of $g_{min} = 0.6g_0$ during
both check intervals,
where $g_0=2\pi \times 16$ MHz is the maximum possible coupling.
By comparing the peak positions of the observed normal-mode splitting
(Fig.~\ref{Figure2}) to Eq.~1, we find that the remaining sample shows an average
atom-cavity coupling of
$g = 0.7 g_0$, exceeding both the cavity field decay rate of
$\kappa = 2\pi\times 1.25$ MHz and the decay rate of the atomic
polarization $\gamma = 2\pi\times 3$ MHz.\par
Such a postselection protocol has already been applied successfully
for measurements of the normal-mode spectrum\cite{Maunz05}.
Only near $\Delta_a \approx \Delta_c = 0$ in Fig.~\ref{Figure3},
the post-selection
protocol is shot-noise limited since even moderate atom-cavity
couplings cause a drop of the transmission to practically zero.
Hence we are unable to filter out all of the less-well localized
atoms, and the data show a systematic increase of the transmitted
intensity for $|\Delta_c| \le 2\pi\times 3$ MHz in Fig.~\ref{Figure3}.
Notice that the simulations, which mimic our experiment,
also reproduce this enhancement.
Remnants in the transmission signal of the pure cavity resonance
have also been reported in a beam of atoms strongly coupled to a
cavity\cite{Childs96}, where it was attributed to fluctuations
in the number of atoms present in the cavity over time,
and for a quantum dot inside a cavity\cite{Hennessy07}
where the increase stems from the fluctuations in emitter energy.\par
Finally, we mention that in Fig.~\ref{Figure2} we did not attempt to
accurately resolve states $|1,+\rangle$ and $|2,+\rangle$ because
this requires extremely long measurement times due to high heating rates
in this region, which makes gathering statistics on strongly coupled
atoms difficult. To give an order of magnitude, 250 hours of
measurement time were required for Fig.~\ref{Figure2}, and more for Fig.~\ref{Figure3}. In
principle, one could imagine reversing the asymmetry in Fig.~\ref{Figure2}, such
that the peaks for state $|1,+\rangle$ and $|2,+\rangle$ are low
while the peaks at $|1,-\rangle$ and $|2,-\rangle$ are enhanced.
However, the check interval which is required in our protocol would
then coincide with a region of cavity-induced heating, given by
($\Delta_c=0,\Delta_a<0$), rendering measurements impossible.
It is for the same reason that we did not attempt to resolve state
$|2,+\rangle$ at $\Delta_c>0$ in Fig.~\ref{Figure3}.\par
\section{Theoretical analysis of the data \label{App::Analysis}}
\subsection{Simulations}
\subsubsection{General information}
We include effects of motion and mismatch between the dipole-trap mode
and the near-resonant mode by numerically simulating
atomic trajectories in the three-dimensional space from injection into
the trap until escape. The measurement procedure is imitated
step by step, and we apply the same evaluation protocol to the resulting
transmission curves.
The motion of the center of mass of the atom is treated classically
in all simulations and obeys a Langevin equation with conservative,
friction and random forces for diffusion processes.
The dipole trap light is modeled by an ac-Stark
shift of the atomic levels.
We have studied three models. Two types of them are based on
the Jaynes-Cummings model for a single atom
strongly coupled to a single cavity mode as described by a master
equation, including dissipation and expanded by a term
accounting for the pumping of the cavity mode. For the first set of
simulations we use
at most one quantum of excitation (where the spectrum is classical),
whereas for the other set we include higher orders to see quantum effects
in the spectrum.
The third kind of simulation is based on the equations for a two-state
atom coupled to a classical field (semiclassical bistability theory).
\subsubsection{Simulations using at most one quantum of excitation}
Here we truncated  the Hilbert space of the atom+mode system after the
first doublet of dressed states.
This approximation allows for an analytical solution
for the forces\cite{Hechenblaikner98} and hence fast computation; it
is appropriate for describing the normal modes in the case of small
driving of the system\cite{Puppe07a}, but inherently excludes any
multiphoton effects. We use it in order to quantify the contribution of
single-photon transitions to the spectra (dotted lines in Fig.~\ref{Figure2} and
Fig.~\ref{Figure3}).
\subsubsection{Simulations using more than one quantum of excitation}
In a second set of simulations we numerically calculate the
excitation of atom and mode as well as the forces on the atom in the
presence of higher doublets. In this way, we are able to
quantitatively reproduce the measurements (dashed lines in Fig.~\ref{Figure2}
and Fig.~\ref{Figure3}) for all spectra.
Since there are no analytical expressions
available for the quantities of interest, we calculate all numerically,
including friction and diffusion tensors which are needed at every
step in the integration of the atomic motion.
For the simulations shown in Fig.~\ref{Figure3}, four Fock states were used.
Simulations with higher Fock states are beyond our computational
resources; as an example, the simulations for Fig.~\ref{Figure3} ran for about 150
hours on 30 nodes of a computer cluster.
We suspect that the absence of multi-photon
transitions of the order four and higher in the code at least
partly explains the deviation between theory and experiment
in the region of the state $|3,-\rangle$.
This is supported by our analysis using the fixed-atom
theory, see last section, where we can easily check the convergence
of the steady-state solutions such as the average intracavity  photon
number. Doing so, we indeed see that a fifth Fock state increases the
average photon number, and this only around the state $|3,-\rangle$.
The convergence around the second dressed state  $|2,-\rangle$ has
already settled.
Further details on the simulations will be shown elsewhere.
\begin{figure}
\begin{center}
\includegraphics[width=\columnwidth]{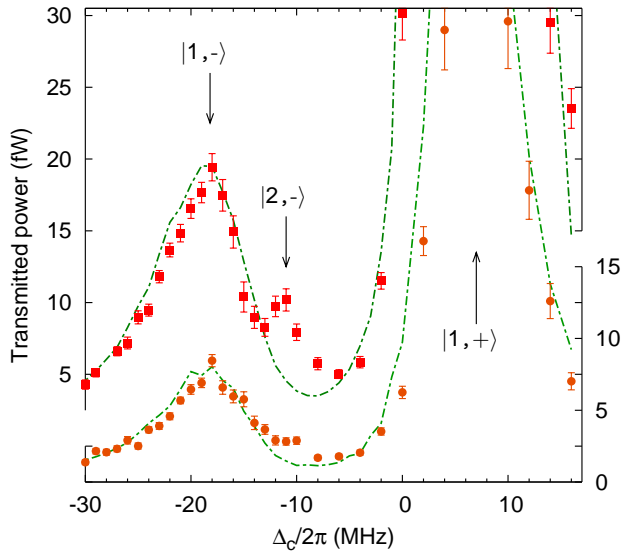}
\caption{The data from Fig.~\ref{Figure2}
  compared to the simulations using the Maxwell-Bloch equations
  (semiclassical optical bistability theory).
  This theory does not reproduce the two-photon
  resonance and remains close to the fully classical theory.
  See Discussion for details.
    \label{fig:supp1}}
\end{center}
\end{figure}
\begin{figure}
\includegraphics[width=\columnwidth]{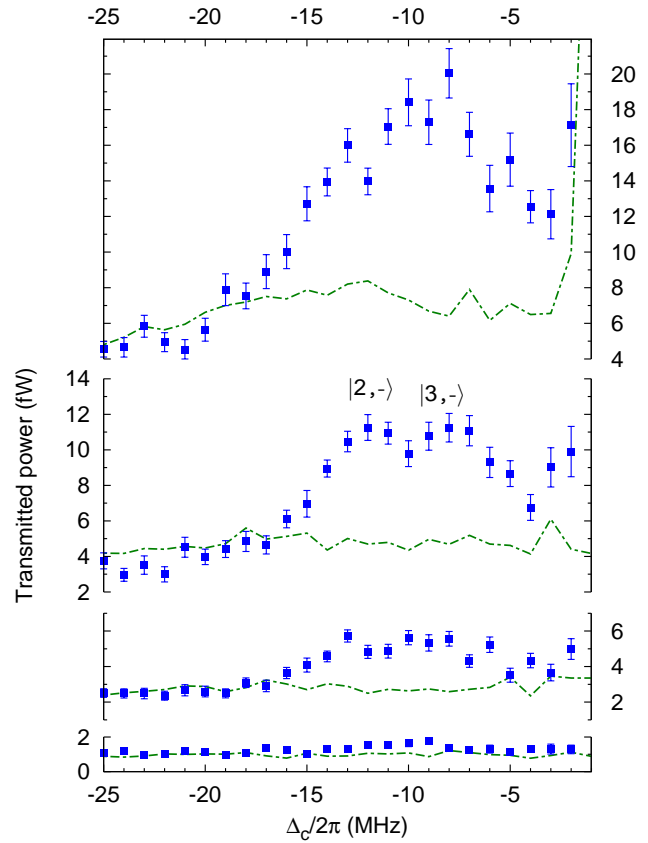}
\caption{ The data from Fig.~\ref{Figure3}
  compared to the simulations using the Maxwell-Bloch equations
  (semiclassical optical bistability theory).
  This theory does not reproduce the two-photon
  resonance and remains close to the fully classical theory.
  See Discussion for details.
    \label{fig:supp2}
}
\end{figure}
\subsubsection{Simulations using the Maxwell-Bloch equations}
In a third set of simulations we considered the {Max\-well-Bloch}
equations, which include the possibility of bistability behavior in
cavity QED. This theory is important because it allows us to quantify
the nonlinearity induced by saturation of the two-state atom
due to its coupling to a classical field.
The results are shown in the  Figs. \ref{fig:supp1} and \ref{fig:supp2}.
\subsection{Analysis of the nonlinear response}
\begin{figure}
\includegraphics[width=\columnwidth]{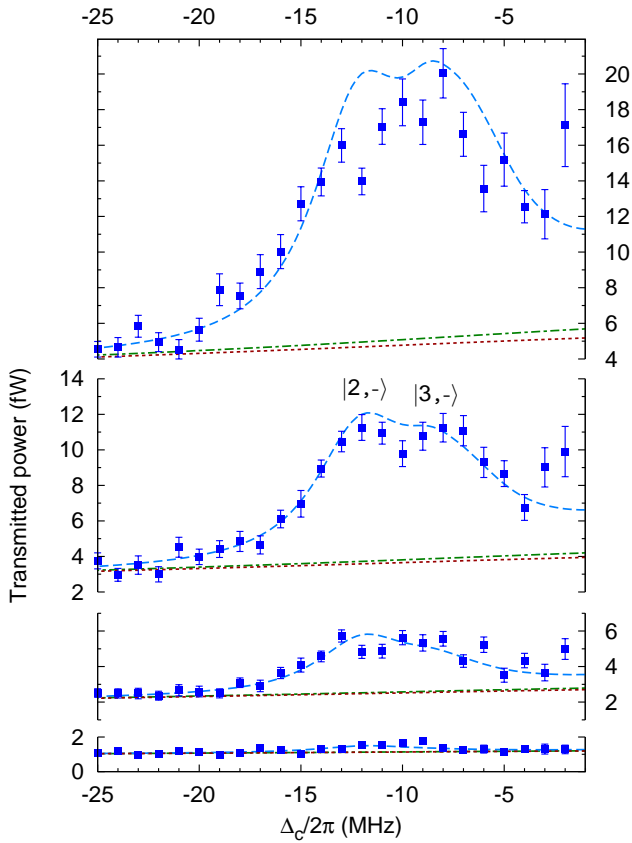}
\caption{The data from Fig.~\ref{Figure3}
compared to the theories for an atom with fixed coupling and
detuning, in the classical
(dotted lines), semiclassical (dash-dotted lines) or quantum
(dashed lines) cases. For fit parameters, see Discussion.
\label{fig:supp3}
}
\end{figure}

In analyzing Fig.~\ref{Figure4} we choose to compare the measured nonlinearity
to the ideal nonlinearity of the driven Jaynes-Cummings model, where
the coupling to the mode is assumed fixed. We start by fitting the
whole series of spectra from Fig.~\ref{Figure3} to the quantum theory. Here, 15
Fock states are used to ensure convergence of the theory, which sets
in at about 5 Fock states. The best match is found for the
parameters $(g,\Delta_a)=2\pi\times(11.5,1)$~MHz, which are fully
consistent with our knowledge of the system. Also notice that the
value of the coupling differs from that obtained from the analysis
from Fig.~\ref{Figure2} by only $0.3$ MHz. We use these parameters as
input for the semiclassical and classical theories. The data and
theories are shown in Fig. \ref{fig:supp3}.\par
To achieve an overall agreement of quantum theory to the data, we need
to shift the theory curves upwards by a small constant offset
$(0.8,1.5,2,2.5)$~fW for the consecutive spectra. For consistency,
the same offset has been applied to all theories. We note two things
about this offset: first, it originates from atomic motion because it
is only needed within the frame of a fixed-atom theory, while the
quantum simulations, which include the motion,
directly match the measurements. Second, this offset is
frequency-independent. Therefore, it is present in the far
off-resonant region, $|\Delta_c|\gg g$, where all theories
asymptotically coincide. Indeed, one can see that the off-resonant
region is already correctly described by the single-photon simulations.
Also notice that the fixed-atom theory needs to
be shifted upwards already for the lowest intensity scan at $0.5$ pW,
where the influence of two-photon transitions on the spectrum is
small. This means that the offset is largely single-photon induced.
We therefore conclude that the offset stems from the motion of
the atom in the trap which leads to a small background transmission
caused by single-photon transitions.
Besides this caveat, the quantum fixed-atom theory reproduces the
observed peak structures for
all four input intensities using one common pair of parameters
$(g,\Delta_a)$. Only in the case of the highest input intensity
($3.3$~pW), the observed peak height does not quite
reach the theory curve. It is plausible to assume that this region is
affected most by heating, since here the light intensity and atomic
excitation reach a maximum, although remaining at a low absolute value
(the excitation probability is below $0.05$ as deduced
from a fixed-atom theory and $0.07$ as deduced from the simulations).
\par
To determine the nonlinearity of the intensity response from the
fixed-atom theory, we proceed in the same way as for the experiment:
for each intensity, we average on the two-photon resonance signal and
off-resonance signal and calculate the difference to obtain the
multi-photon contribution (solid line in Fig.~\ref{Figure4}).
The resulting intensity response shows a mainly quadratic behavior for
the given intensities.
We do the same for the nonlinear response expected from
the Maxwell-Bloch equations (dash-dotted line in Fig.~\ref{fig:supp3},
dashed line in Fig.~\ref{Figure4}).\par
Finally, in Fig.~\ref{fig:supp3} notice the double-peak structure of
the resonance,
which stems from a two-photon and a weak three-photon transition.
For our parameters, the higher multi-photon transitions do not appear
as separate peaks on the right of state $|3,-\rangle$, but do raise the
transmission around this peak.

 \begin{acknowledgments}
We thank N. Syassen for early contributions.
Partial support by the Bavarian PhD programme of excellence QCCC,
the DFG research unit 635, the DFG cluster of excellence MAP and the
EU project SCALA are gratefully acknowledged.
\end{acknowledgments}


\begin{thebibliography}{31}
\expandafter\ifx\csname natexlab\endcsname\relax\def\natexlab#1{#1}\fi
\expandafter\ifx\csname bibnamefont\endcsname\relax
  \def\bibnamefont#1{#1}\fi
\expandafter\ifx\csname bibfnamefont\endcsname\relax
  \def\bibfnamefont#1{#1}\fi
\expandafter\ifx\csname citenamefont\endcsname\relax
  \def\citenamefont#1{#1}\fi
\expandafter\ifx\csname url\endcsname\relax
  \def\url#1{\texttt{#1}}\fi
\expandafter\ifx\csname urlprefix\endcsname\relax\def\urlprefix{URL }\fi
\providecommand{\bibinfo}[2]{#2}
\providecommand{\eprint}[2][]{\url{#2}}

\bibitem[{\citenamefont{Berman}(1994)}]{Berman94}
\bibinfo{editor}{\bibfnamefont{P.~R.} \bibnamefont{Berman}}, ed.,
  \emph{\bibinfo{title}{{Cavity quantum electrodynamics}}}
  (\bibinfo{publisher}{Advances in atomic, molecular, and optical physics, New
  York: Academic Press}, \bibinfo{year}{1994}).

\bibitem[{\citenamefont{Boca et~al.}(2004)\citenamefont{Boca, Miller, Birnbaum,
  Boozer, McKeever, and Kimble}}]{Boca04}
\bibinfo{author}{\bibfnamefont{A.}~\bibnamefont{Boca}},
  \bibinfo{author}{\bibfnamefont{R.}~\bibnamefont{Miller}},
  \bibinfo{author}{\bibfnamefont{K.~M.} \bibnamefont{Birnbaum}},
  \bibinfo{author}{\bibfnamefont{A.~D.} \bibnamefont{Boozer}},
  \bibinfo{author}{\bibfnamefont{J.}~\bibnamefont{McKeever}}, \bibnamefont{and}
  \bibinfo{author}{\bibfnamefont{H.~J.} \bibnamefont{Kimble}},
  \bibinfo{journal}{Phys. Rev. Lett.} \textbf{\bibinfo{volume}{93}},
  \bibinfo{pages}{233603} (\bibinfo{year}{2004}).

\bibitem[{\citenamefont{Maunz et~al.}(2005)\citenamefont{Maunz, Puppe,
  Schuster, Syassen, Pinkse, and Rempe}}]{Maunz05}
\bibinfo{author}{\bibfnamefont{P.}~\bibnamefont{Maunz}},
  \bibinfo{author}{\bibfnamefont{T.}~\bibnamefont{Puppe}},
  \bibinfo{author}{\bibfnamefont{I.}~\bibnamefont{Schuster}},
  \bibinfo{author}{\bibfnamefont{N.}~\bibnamefont{Syassen}},
  \bibinfo{author}{\bibfnamefont{P.~W.~H.} \bibnamefont{Pinkse}},
  \bibnamefont{and} \bibinfo{author}{\bibfnamefont{G.}~\bibnamefont{Rempe}},
  \bibinfo{journal}{Phys. Rev. Lett.} \textbf{\bibinfo{volume}{94}},
  \bibinfo{pages}{033002} (\bibinfo{year}{2005}).

\bibitem[{\citenamefont{Puppe et~al.}(2007{\natexlab{a}})\citenamefont{Puppe,
  Schuster, Grothe, Kubanek, Murr, Pinkse, and Rempe}}]{Puppe07}
\bibinfo{author}{\bibfnamefont{T.}~\bibnamefont{Puppe}},
  \bibinfo{author}{\bibfnamefont{I.}~\bibnamefont{Schuster}},
  \bibinfo{author}{\bibfnamefont{A.}~\bibnamefont{Grothe}},
  \bibinfo{author}{\bibfnamefont{A.}~\bibnamefont{Kubanek}},
  \bibinfo{author}{\bibfnamefont{K.}~\bibnamefont{Murr}},
  \bibinfo{author}{\bibfnamefont{P.~W.~H.} \bibnamefont{Pinkse}},
  \bibnamefont{and} \bibinfo{author}{\bibfnamefont{G.}~\bibnamefont{Rempe}},
  \bibinfo{journal}{Phys. Rev. Lett.} \textbf{\bibinfo{volume}{99}},
  \bibinfo{eid}{013002} (pages~\bibinfo{numpages}{4})
  (\bibinfo{year}{2007}{\natexlab{a}}).

\bibitem[{\citenamefont{{Wallraff} et~al.}(2004)\citenamefont{{Wallraff},
  {Schuster}, {Blais}, {Frunzio}, {Huang}, {Majer}, {Kumar}, {Girvin}, and
  {Schoelkopf}}}]{Wallraff04}
\bibinfo{author}{\bibfnamefont{A.}~\bibnamefont{{Wallraff}}},
  \bibinfo{author}{\bibfnamefont{D.~I.} \bibnamefont{{Schuster}}},
  \bibinfo{author}{\bibfnamefont{A.}~\bibnamefont{{Blais}}},
  \bibinfo{author}{\bibfnamefont{L.}~\bibnamefont{{Frunzio}}},
  \bibinfo{author}{\bibfnamefont{R.-S.} \bibnamefont{{Huang}}},
  \bibinfo{author}{\bibfnamefont{J.}~\bibnamefont{{Majer}}},
  \bibinfo{author}{\bibfnamefont{S.}~\bibnamefont{{Kumar}}},
  \bibinfo{author}{\bibfnamefont{S.~M.} \bibnamefont{{Girvin}}},
  \bibnamefont{and} \bibinfo{author}{\bibfnamefont{R.~J.}
  \bibnamefont{{Schoelkopf}}}, \bibinfo{journal}{Nature}
  \textbf{\bibinfo{volume}{431}}, \bibinfo{pages}{162} (\bibinfo{year}{2004}).

\bibitem[{\citenamefont{{Reithmaier} et~al.}(2004)\citenamefont{{Reithmaier},
  {S{\c e}k}, {L{\"o}ffler}, {Hofmann}, {Kuhn}, {Reitzenstein}, {Keldysh},
  {Kulakovskii}, {Reinecke}, and {Forchel}}}]{Reithmaier04}
\bibinfo{author}{\bibfnamefont{J.~P.} \bibnamefont{{Reithmaier}}},
  \bibinfo{author}{\bibfnamefont{G.}~\bibnamefont{{S{\c e}k}}},
  \bibinfo{author}{\bibfnamefont{A.}~\bibnamefont{{L{\"o}ffler}}},
  \bibinfo{author}{\bibfnamefont{C.}~\bibnamefont{{Hofmann}}},
  \bibinfo{author}{\bibfnamefont{S.}~\bibnamefont{{Kuhn}}},
  \bibinfo{author}{\bibfnamefont{S.}~\bibnamefont{{Reitzenstein}}},
  \bibinfo{author}{\bibfnamefont{L.~V.} \bibnamefont{{Keldysh}}},
  \bibinfo{author}{\bibfnamefont{V.~D.} \bibnamefont{{Kulakovskii}}},
  \bibinfo{author}{\bibfnamefont{T.~L.} \bibnamefont{{Reinecke}}},
  \bibnamefont{and}
  \bibinfo{author}{\bibfnamefont{A.}~\bibnamefont{{Forchel}}},
  \bibinfo{journal}{Nature} \textbf{\bibinfo{volume}{432}},
  \bibinfo{pages}{197} (\bibinfo{year}{2004}).

\bibitem[{\citenamefont{{Yoshie} et~al.}(2004)\citenamefont{{Yoshie},
  {Scherer}, {Hendrickson}, {Khitrova}, {Gibbs}, {Rupper}, {Ell}, {Shchekin},
  and {Deppe}}}]{Yoshie04}
\bibinfo{author}{\bibfnamefont{T.}~\bibnamefont{{Yoshie}}},
  \bibinfo{author}{\bibfnamefont{A.}~\bibnamefont{{Scherer}}},
  \bibinfo{author}{\bibfnamefont{J.}~\bibnamefont{{Hendrickson}}},
  \bibinfo{author}{\bibfnamefont{G.}~\bibnamefont{{Khitrova}}},
  \bibinfo{author}{\bibfnamefont{H.~M.} \bibnamefont{{Gibbs}}},
  \bibinfo{author}{\bibfnamefont{G.}~\bibnamefont{{Rupper}}},
  \bibinfo{author}{\bibfnamefont{C.}~\bibnamefont{{Ell}}},
  \bibinfo{author}{\bibfnamefont{O.~B.} \bibnamefont{{Shchekin}}},
  \bibnamefont{and} \bibinfo{author}{\bibfnamefont{D.~G.}
  \bibnamefont{{Deppe}}}, \bibinfo{journal}{Nature}
  \textbf{\bibinfo{volume}{432}}, \bibinfo{pages}{200} (\bibinfo{year}{2004}).

\bibitem[{\citenamefont{Peter et~al.}(2005)\citenamefont{Peter, Senellart,
  Martrou, Lemaitre, Hours, Gerard, and Bloch}}]{Peter05}
\bibinfo{author}{\bibfnamefont{E.}~\bibnamefont{Peter}},
  \bibinfo{author}{\bibfnamefont{P.}~\bibnamefont{Senellart}},
  \bibinfo{author}{\bibfnamefont{D.}~\bibnamefont{Martrou}},
  \bibinfo{author}{\bibfnamefont{A.}~\bibnamefont{Lemaitre}},
  \bibinfo{author}{\bibfnamefont{J.}~\bibnamefont{Hours}},
  \bibinfo{author}{\bibfnamefont{J.~M.} \bibnamefont{Gerard}},
  \bibnamefont{and} \bibinfo{author}{\bibfnamefont{J.}~\bibnamefont{Bloch}},
  \bibinfo{journal}{Phys. Rev. Lett.} \textbf{\bibinfo{volume}{95}},
  \bibinfo{pages}{067401} (\bibinfo{year}{2005}).

\bibitem[{\citenamefont{{Khitrova} et~al.}(2006)\citenamefont{{Khitrova},
  {Gibbs}, {Kira}, {Koch}, and {Scherer}}}]{Khitrova06}
\bibinfo{author}{\bibfnamefont{G.}~\bibnamefont{{Khitrova}}},
  \bibinfo{author}{\bibfnamefont{H.~M.} \bibnamefont{{Gibbs}}},
  \bibinfo{author}{\bibfnamefont{M.}~\bibnamefont{{Kira}}},
  \bibinfo{author}{\bibfnamefont{S.~W.} \bibnamefont{{Koch}}},
  \bibnamefont{and}
  \bibinfo{author}{\bibfnamefont{A.}~\bibnamefont{{Scherer}}},
  \bibinfo{journal}{Nat. Phys.} \textbf{\bibinfo{volume}{2}},
  \bibinfo{pages}{81} (\bibinfo{year}{2006}).

\bibitem[{\citenamefont{Press et~al.}(2007)\citenamefont{Press, Gotzinger,
  Reitzenstein, Hofmann, Loffler, Kamp, Forchel, and Yamamoto}}]{Press07}
\bibinfo{author}{\bibfnamefont{D.}~\bibnamefont{Press}},
  \bibinfo{author}{\bibfnamefont{S.}~\bibnamefont{Gotzinger}},
  \bibinfo{author}{\bibfnamefont{S.}~\bibnamefont{Reitzenstein}},
  \bibinfo{author}{\bibfnamefont{C.}~\bibnamefont{Hofmann}},
  \bibinfo{author}{\bibfnamefont{A.}~\bibnamefont{Loffler}},
  \bibinfo{author}{\bibfnamefont{M.}~\bibnamefont{Kamp}},
  \bibinfo{author}{\bibfnamefont{A.}~\bibnamefont{Forchel}}, \bibnamefont{and}
  \bibinfo{author}{\bibfnamefont{Y.}~\bibnamefont{Yamamoto}},
  \bibinfo{journal}{Phys. Rev. Lett.} \textbf{\bibinfo{volume}{98}},
  \bibinfo{eid}{117402} (pages~\bibinfo{numpages}{4}) (\bibinfo{year}{2007}).

\bibitem[{\citenamefont{{Hennessy} et~al.}(2007)\citenamefont{{Hennessy},
  {Badolato}, {Winger}, {Gerace}, {Atat{\"u}re}, {Gulde}, {F{\"a}lt}, {Hu}, and
  {Imamo{\u g}lu}}}]{Hennessy07}
\bibinfo{author}{\bibfnamefont{K.}~\bibnamefont{{Hennessy}}},
  \bibinfo{author}{\bibfnamefont{A.}~\bibnamefont{{Badolato}}},
  \bibinfo{author}{\bibfnamefont{M.}~\bibnamefont{{Winger}}},
  \bibinfo{author}{\bibfnamefont{D.}~\bibnamefont{{Gerace}}},
  \bibinfo{author}{\bibfnamefont{M.}~\bibnamefont{{Atat{\"u}re}}},
  \bibinfo{author}{\bibfnamefont{S.}~\bibnamefont{{Gulde}}},
  \bibinfo{author}{\bibfnamefont{S.}~\bibnamefont{{F{\"a}lt}}},
  \bibinfo{author}{\bibfnamefont{E.~L.} \bibnamefont{{Hu}}}, \bibnamefont{and}
  \bibinfo{author}{\bibfnamefont{A.}~\bibnamefont{{Imamo{\u g}lu}}},
  \bibinfo{journal}{Nature} \textbf{\bibinfo{volume}{445}},
  \bibinfo{pages}{896} (\bibinfo{year}{2007}).

\bibitem[{\citenamefont{Carmichael et~al.}(1994)\citenamefont{Carmichael, Tian,
  Ren, and Alsing}}]{Carmichael94}
\bibinfo{author}{\bibfnamefont{H.~J.} \bibnamefont{Carmichael}},
  \bibinfo{author}{\bibfnamefont{L.}~\bibnamefont{Tian}},
  \bibinfo{author}{\bibfnamefont{W.}~\bibnamefont{Ren}}, \bibnamefont{and}
  \bibinfo{author}{\bibfnamefont{P.}~\bibnamefont{Alsing}}, in
  \emph{\bibinfo{booktitle}{{Cavity quantum electrodynamics}}}, edited by
  \bibinfo{editor}{\bibfnamefont{P.~R.} \bibnamefont{Berman}}
  (\bibinfo{publisher}{Advances in atomic, molecular, and optical physics, New
  York: Academic Press}, \bibinfo{year}{1994}), pp. \bibinfo{pages}{381--423}.

\bibitem[{\citenamefont{{Rempe} et~al.}(1987)\citenamefont{{Rempe}, {Walther},
  and {Klein}}}]{Rempe87}
\bibinfo{author}{\bibfnamefont{G.}~\bibnamefont{{Rempe}}},
  \bibinfo{author}{\bibfnamefont{H.}~\bibnamefont{{Walther}}},
  \bibnamefont{and} \bibinfo{author}{\bibfnamefont{N.}~\bibnamefont{{Klein}}},
  \bibinfo{journal}{Phys. Rev. Lett.} \textbf{\bibinfo{volume}{58}},
  \bibinfo{pages}{353} (\bibinfo{year}{1987}).

\bibitem[{\citenamefont{Brune et~al.}(1996)\citenamefont{Brune, Schmidt-Kaler,
  Maali, Dreyer, Hagley, Raimond, and Haroche}}]{Brune96}
\bibinfo{author}{\bibfnamefont{M.}~\bibnamefont{Brune}},
  \bibinfo{author}{\bibfnamefont{F.}~\bibnamefont{Schmidt-Kaler}},
  \bibinfo{author}{\bibfnamefont{A.}~\bibnamefont{Maali}},
  \bibinfo{author}{\bibfnamefont{J.}~\bibnamefont{Dreyer}},
  \bibinfo{author}{\bibfnamefont{E.}~\bibnamefont{Hagley}},
  \bibinfo{author}{\bibfnamefont{J.~M.} \bibnamefont{Raimond}},
  \bibnamefont{and} \bibinfo{author}{\bibfnamefont{S.}~\bibnamefont{Haroche}},
  \bibinfo{journal}{Phys. Rev. Lett.} \textbf{\bibinfo{volume}{76}},
  \bibinfo{pages}{1800} (\bibinfo{year}{1996}).

\bibitem[{\citenamefont{{Schuster} et~al.}(2007)\citenamefont{{Schuster},
  {Houck}, {Schreier}, {Wallraff}, {Gambetta}, {Blais}, {Frunzio}, {Majer},
  {Johnson}, {Devoret} et~al.}}]{Schuster07}
\bibinfo{author}{\bibfnamefont{D.~I.} \bibnamefont{{Schuster}}},
  \bibinfo{author}{\bibfnamefont{A.~A.} \bibnamefont{{Houck}}},
  \bibinfo{author}{\bibfnamefont{J.~A.} \bibnamefont{{Schreier}}},
  \bibinfo{author}{\bibfnamefont{A.}~\bibnamefont{{Wallraff}}},
  \bibinfo{author}{\bibfnamefont{J.~M.} \bibnamefont{{Gambetta}}},
  \bibinfo{author}{\bibfnamefont{A.}~\bibnamefont{{Blais}}},
  \bibinfo{author}{\bibfnamefont{L.}~\bibnamefont{{Frunzio}}},
  \bibinfo{author}{\bibfnamefont{J.}~\bibnamefont{{Majer}}},
  \bibinfo{author}{\bibfnamefont{B.}~\bibnamefont{{Johnson}}},
  \bibinfo{author}{\bibfnamefont{M.~H.} \bibnamefont{{Devoret}}},
  \bibnamefont{et~al.}, \bibinfo{journal}{Nature}
  \textbf{\bibinfo{volume}{445}}, \bibinfo{pages}{515} (\bibinfo{year}{2007}).

\bibitem[{\citenamefont{{Rempe} et~al.}(1991)\citenamefont{{Rempe}, {Thompson},
  {Brecha}, {Lee}, and {Kimble}}}]{Rempe91}
\bibinfo{author}{\bibfnamefont{G.}~\bibnamefont{{Rempe}}},
  \bibinfo{author}{\bibfnamefont{R.~J.} \bibnamefont{{Thompson}}},
  \bibinfo{author}{\bibfnamefont{R.~J.} \bibnamefont{{Brecha}}},
  \bibinfo{author}{\bibfnamefont{W.~D.} \bibnamefont{{Lee}}}, \bibnamefont{and}
  \bibinfo{author}{\bibfnamefont{H.~J.} \bibnamefont{{Kimble}}},
  \bibinfo{journal}{Phys. Rev. Lett.} \textbf{\bibinfo{volume}{67}},
  \bibinfo{pages}{1727} (\bibinfo{year}{1991}).

\bibitem[{\citenamefont{Lugiato and Narducci}(1992)}]{Lugiato92}
\bibinfo{author}{\bibfnamefont{L.~A.} \bibnamefont{Lugiato}} \bibnamefont{and}
  \bibinfo{author}{\bibfnamefont{L.~M.} \bibnamefont{Narducci}}, in
  \emph{\bibinfo{booktitle}{Fundamental Systems in Quantum Optics, Les Houches,
  Session LIII, 1990}}, edited by
  \bibinfo{editor}{\bibfnamefont{J.}~\bibnamefont{Dalibard}},
  \bibinfo{editor}{\bibfnamefont{J.~M.} \bibnamefont{Raimond}},
  \bibnamefont{and}
  \bibinfo{editor}{\bibfnamefont{J.}~\bibnamefont{Zinn-Justin}}
  (\bibinfo{publisher}{Elsevier Science}, \bibinfo{address}{North-Holland,
  Amsterdam}, \bibinfo{year}{1992}), p. \bibinfo{pages}{941}.

\bibitem[{\citenamefont{{Srinivasan} and {Painter}}(2007)}]{Srinivasan07}
\bibinfo{author}{\bibfnamefont{K.}~\bibnamefont{{Srinivasan}}}
  \bibnamefont{and}
  \bibinfo{author}{\bibfnamefont{O.}~\bibnamefont{{Painter}}},
  \bibinfo{journal}{Nature} \textbf{\bibinfo{volume}{450}},
  \bibinfo{pages}{862} (\bibinfo{year}{2007}).

\bibitem[{\citenamefont{{Englund} et~al.}(2007)\citenamefont{{Englund},
  {Faraon}, {Stoltz}, {Petroff}, and {Vu{\v c}kovi{\'c}}}}]{Englund07}
\bibinfo{author}{\bibfnamefont{D.}~\bibnamefont{{Englund}}},
  \bibinfo{author}{\bibfnamefont{A.}~\bibnamefont{{Faraon}}},
  \bibinfo{author}{\bibfnamefont{N.}~\bibnamefont{{Stoltz}}},
  \bibinfo{author}{\bibfnamefont{P.}~\bibnamefont{{Petroff}}},
  \bibnamefont{and} \bibinfo{author}{\bibfnamefont{J.}~\bibnamefont{{Vu{\v
  c}kovi{\'c}}}}, \bibinfo{journal}{Nature} \textbf{\bibinfo{volume}{450}},
  \bibinfo{pages}{857} (\bibinfo{year}{2007}).

\bibitem[{\citenamefont{{Chang} et~al.}(2007)\citenamefont{{Chang},
  {S{\o}rensen}, {Demler}, and {Lukin}}}]{Chang07}
\bibinfo{author}{\bibfnamefont{D.~E.} \bibnamefont{{Chang}}},
  \bibinfo{author}{\bibfnamefont{A.~S.} \bibnamefont{{S{\o}rensen}}},
  \bibinfo{author}{\bibfnamefont{E.~A.} \bibnamefont{{Demler}}},
  \bibnamefont{and} \bibinfo{author}{\bibfnamefont{M.~D.}
  \bibnamefont{{Lukin}}}, \bibinfo{journal}{Nat. Phys.}
  \textbf{\bibinfo{volume}{3}}, \bibinfo{pages}{807} (\bibinfo{year}{2007}).

\bibitem[{\citenamefont{{Carmichael} et~al.}(1996)\citenamefont{{Carmichael},
  {Kochan}, and {Sanders}}}]{Carmichael96}
\bibinfo{author}{\bibfnamefont{H.~J.} \bibnamefont{{Carmichael}}},
  \bibinfo{author}{\bibfnamefont{P.}~\bibnamefont{{Kochan}}}, \bibnamefont{and}
  \bibinfo{author}{\bibfnamefont{B.~C.} \bibnamefont{{Sanders}}},
  \bibinfo{journal}{Phys. Rev. Lett.} \textbf{\bibinfo{volume}{77}},
  \bibinfo{pages}{631} (\bibinfo{year}{1996}).

\bibitem[{\citenamefont{{Thompson} et~al.}(1998)\citenamefont{{Thompson},
  {Turchette}, {Carnal}, and {Kimble}}}]{Thompson98}
\bibinfo{author}{\bibfnamefont{R.~J.} \bibnamefont{{Thompson}}},
  \bibinfo{author}{\bibfnamefont{Q.~A.} \bibnamefont{{Turchette}}},
  \bibinfo{author}{\bibfnamefont{O.}~\bibnamefont{{Carnal}}}, \bibnamefont{and}
  \bibinfo{author}{\bibfnamefont{H.~J.} \bibnamefont{{Kimble}}},
  \bibinfo{journal}{Phys. Rev. A} \textbf{\bibinfo{volume}{57}},
  \bibinfo{pages}{3084} (\bibinfo{year}{1998}).

\bibitem[{\citenamefont{Meekhof et~al.}(1996)\citenamefont{Meekhof, Monroe,
  King, Itano, and Wineland}}]{Meekhof96}
\bibinfo{author}{\bibfnamefont{D.~M.} \bibnamefont{Meekhof}},
  \bibinfo{author}{\bibfnamefont{C.}~\bibnamefont{Monroe}},
  \bibinfo{author}{\bibfnamefont{B.~E.} \bibnamefont{King}},
  \bibinfo{author}{\bibfnamefont{W.~M.} \bibnamefont{Itano}}, \bibnamefont{and}
  \bibinfo{author}{\bibfnamefont{D.~J.} \bibnamefont{Wineland}},
  \bibinfo{journal}{Phys. Rev. Lett.} \textbf{\bibinfo{volume}{76}},
  \bibinfo{pages}{1796} (\bibinfo{year}{1996}).

\bibitem[{\citenamefont{{Mielke} et~al.}(1998)\citenamefont{{Mielke}, {Foster},
  and {Orozco}}}]{Mielke98}
\bibinfo{author}{\bibfnamefont{S.~L.} \bibnamefont{{Mielke}}},
  \bibinfo{author}{\bibfnamefont{G.~T.} \bibnamefont{{Foster}}},
  \bibnamefont{and} \bibinfo{author}{\bibfnamefont{L.~A.}
  \bibnamefont{{Orozco}}}, \bibinfo{journal}{Phys. Rev. Lett.}
  \textbf{\bibinfo{volume}{80}}, \bibinfo{pages}{3948} (\bibinfo{year}{1998}).

\bibitem[{\citenamefont{{Foster} et~al.}(2000)\citenamefont{{Foster}, {Orozco},
  {Castro-Beltran}, and {Carmichael}}}]{Foster00}
\bibinfo{author}{\bibfnamefont{G.~T.} \bibnamefont{{Foster}}},
  \bibinfo{author}{\bibfnamefont{L.~A.} \bibnamefont{{Orozco}}},
  \bibinfo{author}{\bibfnamefont{H.~M.} \bibnamefont{{Castro-Beltran}}},
  \bibnamefont{and} \bibinfo{author}{\bibfnamefont{H.~J.}
  \bibnamefont{{Carmichael}}}, \bibinfo{journal}{Phys. Rev. Lett.}
  \textbf{\bibinfo{volume}{85}}, \bibinfo{pages}{3149} (\bibinfo{year}{2000}).

\bibitem[{\citenamefont{{Birnbaum} et~al.}(2005)\citenamefont{{Birnbaum},
  {Boca}, {Miller}, {Boozer}, {Northup}, and {Kimble}}}]{Birnbaum05}
\bibinfo{author}{\bibfnamefont{K.~M.} \bibnamefont{{Birnbaum}}},
  \bibinfo{author}{\bibfnamefont{A.}~\bibnamefont{{Boca}}},
  \bibinfo{author}{\bibfnamefont{R.}~\bibnamefont{{Miller}}},
  \bibinfo{author}{\bibfnamefont{A.~D.} \bibnamefont{{Boozer}}},
  \bibinfo{author}{\bibfnamefont{T.~E.} \bibnamefont{{Northup}}},
  \bibnamefont{and} \bibinfo{author}{\bibfnamefont{H.~J.}
  \bibnamefont{{Kimble}}}, \bibinfo{journal}{Nature}
  \textbf{\bibinfo{volume}{436}}, \bibinfo{pages}{87} (\bibinfo{year}{2005}).

\bibitem[{\citenamefont{{Nu{\ss}mann} et~al.}(2005)\citenamefont{{Nu{\ss}mann},
  {Murr}, {Hijlkema}, {Weber}, {Kuhn}, and {Rempe}}}]{Nussmann05}
\bibinfo{author}{\bibfnamefont{S.}~\bibnamefont{{Nu{\ss}mann}}},
  \bibinfo{author}{\bibfnamefont{K.}~\bibnamefont{{Murr}}},
  \bibinfo{author}{\bibfnamefont{M.}~\bibnamefont{{Hijlkema}}},
  \bibinfo{author}{\bibfnamefont{B.}~\bibnamefont{{Weber}}},
  \bibinfo{author}{\bibfnamefont{A.}~\bibnamefont{{Kuhn}}}, \bibnamefont{and}
  \bibinfo{author}{\bibfnamefont{G.}~\bibnamefont{{Rempe}}},
  \bibinfo{journal}{Nature Phys.} \textbf{\bibinfo{volume}{1}},
  \bibinfo{pages}{122} (\bibinfo{year}{2005}).

\bibitem[{\citenamefont{Maunz et~al.}(2004)\citenamefont{Maunz, Puppe,
  Schuster, Syassen, Pinkse, and Rempe}}]{Maunz04}
\bibinfo{author}{\bibfnamefont{P.}~\bibnamefont{Maunz}},
  \bibinfo{author}{\bibfnamefont{T.}~\bibnamefont{Puppe}},
  \bibinfo{author}{\bibfnamefont{I.}~\bibnamefont{Schuster}},
  \bibinfo{author}{\bibfnamefont{N.}~\bibnamefont{Syassen}},
  \bibinfo{author}{\bibfnamefont{P.~W.~H.} \bibnamefont{Pinkse}},
  \bibnamefont{and} \bibinfo{author}{\bibfnamefont{G.}~\bibnamefont{Rempe}},
  \bibinfo{journal}{Nature (London)} \textbf{\bibinfo{volume}{428}},
  \bibinfo{pages}{50} (\bibinfo{year}{2004}).

\bibitem[{\citenamefont{Hechenblaikner
  et~al.}(1998)\citenamefont{Hechenblaikner, Gangl, Horak, and
  Ritsch}}]{Hechenblaikner98}
\bibinfo{author}{\bibfnamefont{G.}~\bibnamefont{Hechenblaikner}},
  \bibinfo{author}{\bibfnamefont{M.}~\bibnamefont{Gangl}},
  \bibinfo{author}{\bibfnamefont{P.}~\bibnamefont{Horak}}, \bibnamefont{and}
  \bibinfo{author}{\bibfnamefont{H.}~\bibnamefont{Ritsch}},
  \bibinfo{journal}{Phys. Rev. A} \textbf{\bibinfo{volume}{58}},
  \bibinfo{pages}{3030} (\bibinfo{year}{1998}).

\bibitem[{\citenamefont{Childs et~al.}(1996)\citenamefont{Childs, An, Otteson,
  Dasari, and Feld}}]{Childs96}
\bibinfo{author}{\bibfnamefont{J.~J.} \bibnamefont{Childs}},
  \bibinfo{author}{\bibfnamefont{K.}~\bibnamefont{An}},
  \bibinfo{author}{\bibfnamefont{M.~S.} \bibnamefont{Otteson}},
  \bibinfo{author}{\bibfnamefont{R.~R.} \bibnamefont{Dasari}},
  \bibnamefont{and} \bibinfo{author}{\bibfnamefont{M.~S.} \bibnamefont{Feld}},
  \bibinfo{journal}{Phys. Rev. Lett.} \textbf{\bibinfo{volume}{77}},
  \bibinfo{pages}{2901} (\bibinfo{year}{1996}).

\bibitem[{\citenamefont{Puppe et~al.}(2007{\natexlab{b}})\citenamefont{Puppe,
  Schuster, Maunz, Murr, Pinkse, and Rempe}}]{Puppe07a}
\bibinfo{author}{\bibfnamefont{T.}~\bibnamefont{Puppe}},
  \bibinfo{author}{\bibfnamefont{I.}~\bibnamefont{Schuster}},
  \bibinfo{author}{\bibfnamefont{P.}~\bibnamefont{Maunz}},
  \bibinfo{author}{\bibfnamefont{K.}~\bibnamefont{Murr}},
  \bibinfo{author}{\bibfnamefont{P.~W.~H.} \bibnamefont{Pinkse}},
  \bibnamefont{and} \bibinfo{author}{\bibfnamefont{G.}~\bibnamefont{Rempe}},
  \bibinfo{journal}{J. Mod. Opt.} \textbf{\bibinfo{volume}{54}},
  \bibinfo{pages}{1927} (\bibinfo{year}{2007}{\natexlab{b}}).

\end{thebibliography}

\end{document}